\documentstyle[aps]{revtex}

\newcommand{\bel}[1]{\begin{equation}\label{#1}}
\newcommand{\be}{\begin{equation}}
\newcommand{\ee}{\end{equation}}

\newcommand{\beal}[1]{\begin{eqnarray}\label{#1}}
\newcommand{\bea}{\begin{eqnarray}}
\newcommand{\eea}{\end{eqnarray}}

\newcommand{\bean}{\begin{eqnarray*}}
\newcommand{\eean}{\end{eqnarray*}}

\newcommand{\ba}{\begin{array}}
\newcommand{\ea}{\end{array}}

\newcommand{\bab}{\begin{abstract}}
\newcommand{\eab}{\end{abstract}}

\newcommand{\bml}{\begin{mathletters}}
\newcommand{\eml}{\end{mathletters}}

\newcommand{\q}{\quad}
\newcommand{\qq}{\quad\quad}

\newcommand{\bfm}[1]{\mbox{\boldmath $#1$}}

\newcommand{\dv}{\partial}

\newcommand{\bam}{\left( \begin{array}}
\newcommand{\eam}{\end{array} \right)}

\newcommand{\bamq}[4]{\left( \begin{array}{cccc}{#1}&{#2}&{#3}&{#4}\\}
\newcommand{\bamc}[5]{\left( \begin{array}{ccccc}{#1}&{#2}&{#3}&{#4}&{#5}\\}

\newcommand{\law}{\leftarrow}

\newcommand{\raw}{\rightarrow}
\newcommand{\lraw}{\longrightarrow}

\newcommand{\ag}{\alpha}
\newcommand{\bg}{\beta}

\newcommand{\dg}{\delta}
\newcommand{\eg}{\epsilon}

\newcommand{\pg}{\phi}

\def\xc{{\mathchoice {\setbox0=\hbox{$\displaystyle\rm C$}\hbox{\hbox
to0pt{\kern0.4\wd0\vrule height0.9\ht0\hss}\box0}}
{\setbox0=\hbox{$\textstyle\rm C$}\hbox{\hbox
to0pt{\kern0.4\wd0\vrule height0.9\ht0\hss}\box0}}
{\setbox0=\hbox{$\scriptstyle\rm C$}\hbox{\hbox
to0pt{\kern0.4\wd0\vrule height0.9\ht0\hss}\box0}}
{\setbox0=\hbox{$\scriptscriptstyle\rm C$}\hbox{\hbox
to0pt{\kern0.4\wd0\vrule height0.9\ht0\hss}\box0}}}}

\def\xg{{\mathchoice {\setbox0=\hbox{$\displaystyle\rm G$}\hbox{\hbox
to0pt{\kern0.4\wd0\vrule height0.9\ht0\hss}\box0}}
{\setbox0=\hbox{$\textstyle\rm G$}\hbox{\hbox
to0pt{\kern0.4\wd0\vrule height0.9\ht0\hss}\box0}}
{\setbox0=\hbox{$\scriptstyle\rm G$}\hbox{\hbox
to0pt{\kern0.4\wd0\vrule height0.9\ht0\hss}\box0}}
{\setbox0=\hbox{$\scriptscriptstyle\rm G$}\hbox{\hbox
to0pt{\kern0.4\wd0\vrule height0.9\ht0\hss}\box0}}}}

\def\xi{{\rm I\!I}}

\def\xo{{\mathchoice {\setbox0=\hbox{$\displaystyle\rm O$}\hbox{\hbox
to0pt{\kern0.4\wd0\vrule height0.9\ht0\hss}\box0}}
{\setbox0=\hbox{$\textstyle\rm O$}\hbox{\hbox
to0pt{\kern0.4\wd0\vrule height0.9\ht0\hss}\box0}}
{\setbox0=\hbox{$\scriptstyle\rm O$}\hbox{\hbox
to0pt{\kern0.4\wd0\vrule height0.9\ht0\hss}\box0}}
{\setbox0=\hbox{$\scriptscriptstyle\rm O$}\hbox{\hbox
to0pt{\kern0.4\wd0\vrule height0.9\ht0\hss}\box0}}}}

\def\xq{{\mathchoice {\setbox0=\hbox{$\displaystyle\rm
Q$}\hbox{\raise
0.15\ht0\hbox to0pt{\kern0.4\wd0\vrule height0.8\ht0\hss}\box0}}
{\setbox0=\hbox{$\textstyle\rm Q$}\hbox{\raise
0.15\ht0\hbox to0pt{\kern0.4\wd0\vrule height0.8\ht0\hss}\box0}}
{\setbox0=\hbox{$\scriptstyle\rm Q$}\hbox{\raise
0.15\ht0\hbox to0pt{\kern0.4\wd0\vrule height0.7\ht0\hss}\box0}}
{\setbox0=\hbox{$\scriptscriptstyle\rm Q$}\hbox{\raise
0.15\ht0\hbox to0pt{\kern0.4\wd0\vrule height0.7\ht0\hss}\box0}}}}

\def\xs{{\mathchoice
{\setbox0=\hbox{$\displaystyle     \rm S$}\hbox{\raise0.5\ht0\hbox
to0pt{\kern0.35\wd0\vrule height0.45\ht0\hss}\hbox
to0pt{\kern0.55\wd0\vrule height0.5\ht0\hss}\box0}}
{\setbox0=\hbox{$\textstyle        \rm S$}\hbox{\raise0.5\ht0\hbox
to0pt{\kern0.35\wd0\vrule height0.45\ht0\hss}\hbox
to0pt{\kern0.55\wd0\vrule height0.5\ht0\hss}\box0}}
{\setbox0=\hbox{$\scriptstyle      \rm S$}\hbox{\raise0.5\ht0\hbox
to0pt{\kern0.35\wd0\vrule height0.45\ht0\hss}\raise0.05\ht0\hbox
to0pt{\kern0.5\wd0\vrule height0.45\ht0\hss}\box0}}
{\setbox0=\hbox{$\scriptscriptstyle\rm S$}\hbox{\raise0.5\ht0\hbox
to0pt{\kern0.4\wd0\vrule height0.45\ht0\hss}\raise0.05\ht0\hbox
to0pt{\kern0.55\wd0\vrule height0.45\ht0\hss}\box0}}}}

\def\xt{{\mathchoice {\setbox0=\hbox{$\displaystyle\rm
T$}\hbox{\hbox to0pt{\kern0.3\wd0\vrule height0.9\ht0\hss}\box0}}
{\setbox0=\hbox{$\textstyle\rm T$}\hbox{\hbox
to0pt{\kern0.3\wd0\vrule height0.9\ht0\hss}\box0}}
{\setbox0=\hbox{$\scriptstyle\rm T$}\hbox{\hbox
to0pt{\kern0.3\wd0\vrule height0.9\ht0\hss}\box0}}
{\setbox0=\hbox{$\scriptscriptstyle\rm T$}\hbox{\hbox
to0pt{\kern0.3\wd0\vrule height0.9\ht0\hss}\box0}}}}

\def\xz{{\mathchoice {\hbox{$\sf\textstyle Z\kern-0.4em Z$}}
{\hbox{$\sf\textstyle Z\kern-0.4em Z$}}
{\hbox{$\sf\scriptstyle Z\kern-0.3em Z$}}
{\hbox{$\sf\scriptscriptstyle Z\kern-0.2em Z$}}}}

\newcommand{\fs}{\footnotesize}

\newcommand{\bii}{\begin{itemize}}
\newcommand{\eii}{\end{itemize}}

\newcommand{\ben}{\begin{enumerate}}
\newcommand{\een}{\end{enumerate}}

\newcommand{\bq}{\begin{quote}}
\newcommand{\eq}{\end{quote}}

\newcommand{\bc}{\begin{center}}
\newcommand{\ec}{\end{center}}

\newcommand{\btb}{\begin{table}}
\newcommand{\etb}{\end{table}}

\newcommand{\bt}{\begin{tabular}}
\newcommand{\et}{\end{tabular}}

\newcommand{\br}{\begin{flushright}}
\newcommand{\er}{\end{flushright}}

\newcommand{\bl}{\begin{flushleft}}
\newcommand{\el}{\end{flushleft}}

\newcommand{\bref}{\begin{references}}
\newcommand{\eref}{\end{references}}

\newcommand{\bb}{\begin{thebibliography}{99}}
\newcommand{\eb}{\end{thebibliography}}
\newcommand{\bi}{\bibitem}

\newcommand{\btp}{\begin{titlepage}}
\newcommand{\etp}{\end{titlepage}}

\newcommand{\go}{\section{Introduction}}
\newcommand{\con}{\section{Conclusions}}
\newcommand{\ack}{\section*{Acknowledgments}}

\newcommand{\axp}[3]{Ann.~Phys.~(NY)                    {\bf #1},  #2  (19#3)}

\newcommand{\ixa}[3]{Int.~J.~Mod.~Phys.~A               {\bf #1},  #2  (19#3)}

\newcommand{\jxe}[3]{J.~Math.~Phys.                     {\bf #1},  #2  (19#3)} 

\newcommand{\jxg}[3]{J.~Phys.~A                         {\bf #1},  #2  (19#3)}

\newcommand{\nxb}[3]{Nucl.~Phys.                        {\bf #1},  #2  (19#3)}

\newcommand{\nxd}[3]{Nuovo Cimento                      {\bf #1},  #2  (19#3)}

\newcommand{\pxf}[3]{Phys.~Rev.~D                       {\bf #1},  #2  (19#3)}

\newcommand{\pxi}[3]{Phys.~Lett.                        {\bf #1},  #2  (19#3)}

\newcommand{\pxxa}[3]{Prog.~Theor.~Phys.                {\bf #1},  #2  (19#3)}

\newcommand{\co}{\mbox{\boldmath $\cal C$}}
\newcommand{\qu}{\mbox{\boldmath $\cal H$}}
\newcommand{\oct}{\mbox{\boldmath $\cal O$}}
\newcommand{\rea}{\mbox{\boldmath $\cal R$}}

\title{OCTONIONIC QUANTUM MECHANICS\\ AND\\ COMPLEX GEOMETRY}

\author{Stefano De Leo\thanks{{\sl deleos@le.infn.it}}$^{(a,b)}$ and 
Khaled Abdel-Khalek\thanks{{\sl khaled@le.infn.it}}$^{(a)}$} 
\address{$^{(a)}$~Dipartimento di Fisica - Universit\`a di Lecce\\
$^{(b)}$~Istituto Nazionale di Fisica Nucleare - Sezione di Lecce\\
- Lecce, 73100, Italy -}

\date{Revised Version, July 1996}

\draft

\begin{document}

\maketitle

\bab
The use of  complex geometry allows us to obtain a consistent formulation of 
octonionic quantum mechanics (OQM). In our octonionic formulation we 
solve 
the hermiticity problem and define an appropriate momentum operator within 
OQM. The nonextendability of the completeness relation and the norm 
conservation is also discussed in details.
\eab

\renewcommand{\thefootnote}{\sharp\arabic{footnote}}

\go

In the early thirties, in order to explain the novel 
phenomena of that time, namely $\beta$--decay
 and the strong interactions, Jordan~\cite{jor} introduced a 
nonassociative but commutative algebra as a basic block for a new quantum
theory.  With the discovery that 
$3 \times 3$ hermitian octonionic matrices realize the Jordan 
postulate~\cite{wig,alb1}, octonions appeared, 
for the first time, in quantum mechanics.
The hope of applying nonassociative algebras to physics was soon dashed 
with the Fermi 
theory of the $\beta$--decay and with the Yukawa model of nuclear force. 
Octonions disappeared from physics as soon after being introduced. 

During the early sixties, Finkelstein {\em et al.}~\cite{fink1,fink2} tried to 
extend the standard quantum mechanics by using the quaternionic field. The 
main difficulty, within their framework, was the non existence of proper 
tensor products for quaternionic states. Then, with the quark revolution, 
new problems, like quark confinement, appeared. G\"ursey and 
G\"unaydin~\cite{gur2} tried to explain this problem on a fundamental 
level by investigating the possibility of constructing a consistent 
octonionic quantum mechanics.

An important step towards a  generalization of 
 standard quantum theories is the use 
of complex scalar products~\cite{hor} (or complex geometry as called by 
Rembieli\'nski~\cite{rem}). Without it we cannot define a 
consistent tensor 
product. 

Nonassociative numbers are difficult to manipulate and so the use of the 
octonionic field within OQM is non-trivial. Obviously, if we are not 
being able to 
construct a suitable OQM, octonions will remain beautiful ghosts in search of 
a physical incarnation.

In this work, we overcome the problems due to the nonassociativity of the 
octonionic algebra. Both the quantum mechanics postulates and the octonions 
nonassociativity property will be respected. 

Is there an acceptable generic octonionic quantum theory? Do octonionic quantum 
theories necessitate complex geometry? At this stage these questions lack 
answers and the aim of our work is to clarify these points. 

This article is organized as follows: In section II, we give a brief 
introduction to the octonionic division algebra and 
introduce barred octonions. 
After this mathematical discussion, in section III, we show how the complex 
geometry, the main tool to obtain a suitable formulation of 
OQM, allows us to overcome the hermiticity problem. In this section we 
also introduce the appropriate definition for the momentum operator. In 
section IV, we discuss the octonionic Hilbert space and 
disprove the 
standard objections concerning the nonextendability of completeness relation 
and norm conservation to octonionic quantum mechanics. The future 
developments are drawn in the final section.

\section{Octonionic algebra and barred octonions}

A classical theorem~\cite{sch} states that the only 
division algebra over the reals are algebras of dimensions 1, 2, 4 and 
8, the only associative division algebras over the reals are $\rea$, $\co$ 
and $\qu$ (quaternions), whereas the  {\tt nonassociative} 
algebras include the octonions 
$\oct$ (an interesting discussion concerning nonassociative algebras is 
presented in~\cite{oku}). In this paper we will deal with octonions and 
their generalizations.

We summarize our notation for the octonionic algebra and 
introduce the concept of barred operators. 
There is a number of equivalent ways to represent the octonions 
multiplication table. Fortunately, it is always possible to choose an 
orthonormal basis $(e_0 , \ldots ,e_7 )$ such that
\be
{\cal O}=r_{0}+\sum_{m=1}^{7} r_{m}e_{m} \qq (~r_{0,...,7}~~ \mbox{reals}~) \q , 
\ee 
where $e_{m}$ are elements obeying the noncommutative and nonassociative 
algebra
\be
e_{m}e_{n}=-\dg_{mn}+ \eg_{mnp}e_{p} \qq 
(~\mbox{{\fs $m, \; n, \; p =1,..., 7$}}~) 
\q ,
\ee
with $\eg_{mnp}$ totally antisymmetric and equal to unity for the seven 
combinations 
\[
123, \; 145, \; 176, \; 246, \; 257, \; 347 \; \mbox{and} \; 365  
\]
(each cycle represents a 
quaternionic subalgebra). The octonionic conjugate 
${\cal O}^{\dag}$ is given by
\be
{\cal O}^{\dag}=r_{0}-\sum_{m=1}^{7} r_{m}e_{m} \q . 
\ee 

Let us now introduce the concept of {\tt barred octonions}. Working with 
noncommutative algebras, we  
must distinguish between left and right multiplication
\[ e_m ~ {\cal O} \neq {\cal O} ~ e_m \q . \]
It is appropriate to indicate the left-action of the octonionic 
imaginary units
by
\be e_m \q , \ee
and the right-action by
\be 1 \mid e_m  \qq ~[~(1 \mid e_m) ~ {\cal O} \equiv {\cal O} ~ e_m ~]~ \q .\ee

In recent papers~\cite{deleo1,deleo2} the successful applications of 
barred quaternions in quantum mechanics, field theory and gauge 
theories suggest us to investigate possible potentialities of octonionic 
barred numbers.

\section{Octonionic momentum operator}

We begin this section by presenting an apparently hopeless problem related 
to the nonassociativity of the octonionic field. Working in quantum 
mechanics we require that an antihermitian operator satisfies the following 
relation
\be
\int d{\bf  x}~ \psi^{\dag} (A\pg)= -\int d{\bf  x}~ (A\psi)^{\dag} \pg \q .
\ee
In octonionic quantum mechanics (OQM) we can immediately verify that 
\bfm{\dv} represents an antihermitian operator with all the properties of a 
translation operator. Nevertheless, while in complex (CQM) and quaternionic 
(QQM) quantum mechanics we can define a corresponding hermitian operator 
multiplying by an imaginary unit the operator \bfm{\dv}, one encounters in 
OQM the following problem:
\bc
{\tt no imaginary unit,} \bfm{e_m} {\tt , represents an antihermitian operator \q .}
\ec
In fact, the nonassociativity of the octonionic algebra implies, in general 
(for arbitrary $\psi$ and $\pg$)
\be
\int d{\bf  x}~ \psi^{\dag} (e_{m}\pg) \neq 
-\int d{\bf  x}~ (e_{m}\psi)^{\dag} \pg =
\int d{\bf  x}~ (\psi^{\dag} e_{m}) \pg \qq \mbox{{\fs $(~m=1, ...,7~)$}} \q .
\ee
This contrasts with the situation within complex and quaternionic 
quantum mechanics. Such a difficulty is overcome by using 
complex projection of the scalar product (complex geometry), with respect to 
one of our imaginary 
units. We break the symmetry between the seven imaginary units 
$e_{1}$, ... , $e_{7}$ and choose as projection plane that one characterized by 
$(1, \; e_{1})$. The new scalar product is quickly obtained performing, in 
the standard definition, the following substitution
\[ 
\int d{\bf x} \lraw  \int_{c} d{\bf x} \equiv 
\frac{1-e_{1}\mid e_{1}}{2}~ \int d{\bf x} \q .
\]
Working in OQM with {\tt complex geometry}, 
$e_{1}$ represents an antihermitian operator. In order to simplify the 
proof we write the octonionic functions $\psi$ and $\pg$ as follows:
\bean
\psi & ~=~ & \psi_{1} + e_{2} \psi_{2} + e_{4} \psi_{3} + e_{6} \psi_{4} \q ,\\
\pg & ~=~ & \pg_{1} + e_{2} \pg_{2} + e_{4} \pg_{3} + e_{6} \pg_{4} \q ,\\
    & & ~~~~~~~~~~~[~\psi_{1, ..., 4}~\mbox{and}~\pg_{1, ..., 4} 
          \in \bfm{\cal C}(1, \; e_{1})~] \q .
\eean
The antihermiticity of $e_{1}$ is shown if
\be
\int_{c} d{\bf  x}~ \psi^{\dag} (e_{1}\pg) = 
-\int_{c} d{\bf  x}~ (e_{1}\psi)^{\dag} \pg \q .
\ee
In the previous equation the only nonvanishing terms are represented by 
{\tt diagonal} terms ($\sim \psi_{1}^{\dag}\pg_{1}, 
\; \psi_{2}^{\dag}\pg_{2}, \; \psi_{3}^{\dag}\pg_{3}, 
\; \psi_{4}^{\dag}\pg_{4}$). In fact, {\tt off-diagonal} terms, like 
$\psi_{2}^{\dag}\pg_{3}, \; \psi_{3}^{\dag}\pg_{4}$, are killed by the 
complex projection,
\bean
~(\psi_{2}^{\dag} e_{2})[e_{1}(e_{4}\pg_{3})] & ~\sim ~ & 
(\ag_{2}e_{2}+\ag_{3}e_{3})(\ag_{4}e_{4}+\ag_{5}e_{5}) \sim  
\ag_{6}e_{6} + \ag_{7} e_{7} \q , \\ 
~[(\psi_{3}^{\dag} e_{4}) e_{1}](e_{6}\pg_{4}) & ~\sim ~& 
~(\bg_{4}e_{4}+\bg_{5}e_{5})(\bg_{6}e_{6}+\bg_{7}e_{7}) \sim  
\bg_{2}e_{2} + \bg_{3} e_{3} \q , \\
 & & ~~~~~~~~~~~[~\ag_{2, ..., 7}~\mbox{and}~\bg_{2, ..., 7} \in  
     \bfm{\cal R}~] \q .
\eean
The diagonal terms give
\bml
\beal{dia}
\int_{c} d{\bf  x}~ \psi^{\dag} (e_{1}\pg) & = &
\psi_{1}^{\dag}(e_{1}\pg_{1})
-(\psi_{2}^{\dag}e_{2})[e_{1}(e_{2}\pg_{2})]
-(\psi_{3}^{\dag}e_{4})[e_{1}(e_{4}\pg_{3})]
-(\psi_{4}^{\dag}e_{6})[e_{1}(e_{6}\pg_{4})] \q ,\\
-\int_{c} d{\bf  x}~ (e_{1}\psi)^{\dag} \pg  & = &
(\psi_{1}^{\dag}e_{1})\pg_{1}
-[(\psi_{2}^{\dag}e_{2})e_{1}](e_{2}\pg_{2})
-[(\psi_{3}^{\dag}e_{4})e_{1}](e_{4}\pg_{3})
-[(\psi_{4}^{\dag}e_{6})e_{1}](e_{6}\pg_{4}) \q .
\eea
\eml
The parenthesis in~(\ref{dia}-b) are not of relevance since
\bc
\bt{lll}
$\psi_{1}^{\dag}e_{1}\pg_{1}$ & {\fs ~~~$(1, \; e_{1})$} & ~~~is a complex 
number , \\ 
$\psi_{2}^{\dag} e_{2} e_{1} e_{2}\pg_{2}$ & {\fs ~~~(subalgebra 123)} , & \\ 
$\psi_{3}^{\dag} e_{4} e_{1} e_{4}\pg_{3}$ & {\fs ~~~(subalgebra 145)} , & \\
$\psi_{4}^{\dag} e_{6} e_{1} e_{6}\pg_{4}$ & {\fs ~~~(subalgebra 176)}   & 
are quaternionic numbers .
\et
\ec
The above-mentioned demonstration does not work for the imaginary units 
$e_{2}$, ... , $e_{7}$ (breaking the symmetry between the seven 
octonionic imaginary units). 

Now, we can define an hermitian operator multiplying by $e_{1}$ 
the operator \bfm{\dv}. 
However, such an operator is not expected to commute with the Hamiltonian, 
which will be, in general, an octonionic quantity. The final step towards an 
appropriate definition of the momentum operator is represented by choosing 
as imaginary unit the barred operator $1\mid e_{1}$ (the 
antihermiticity proof is very similar to the previous one). In OQM with 
complex geometry the {\tt the appropriate} momentum operator is then given by
\be 
{\bf p} \equiv - \bfm{\dv} \mid e_{1} \q .
\ee
Obviously, in order to write equations relativistically covariant, we must 
treat the space components and time in the same way, hence we are obliged 
to modify the standard QM operator, $i\dv_{t}$, by the following substitution
\[ i\dv_{t} \lraw \dv_{t} \mid e_{1} \q ,\]
and so the octonionic Dirac equation becomes
\be
\dv_{t} \psi e_{1} = \bfm{\ag} \cdot ({\bf p}\psi)+ m \bg \psi 
\qq (~{\bf p} \equiv - \bfm{\dv} \mid e_{1}~) \q .
\ee
The possibility to write a consistent momentum operator represents for us 
an impressive argument in favor of the use of a complex geometry in 
formulating OQM. Besides, such a complex geometry gives us a welcome 
{\tt quadrupling} of solutions. In fact, 
\[ \psi, \; e_{2}\psi, \; e_{4}\psi, \; e_{6}\psi \qq
\psi \in \bfm{\cal C}(1, \; e_{1}) \]
represent now complex-orthogonal solutions. Therefore, we have the 
possibility to write a one-component octonionic Dirac equation in which all 
four standard Dirac free-particle solutions appear~\cite{dekh}.

\section{Octonionic Hilbert space}

In the early seventies, G\"ursey and G\"unaydin~\cite{gur2}  
have made an attempt to extend
the underlying number field of quantum mechanics 
from complex numbers to octonions. 
The main results are the following: 
One can realize the representations of the Poincar\'e  
group over an octonionic Hilbert space with complex scalar
product which has an automorphism group SU(3). Such 
octonionic Hilbert space can be divided into an 
observable subspace corresponding to the usual complex 
Hilbert space (CHS) of quantum mechanics and 
an unobservable subspace corresponding to
the nonassociative components of the underlying octonionic algebra.

Rembieli\'nski~\cite{rem} established the 
isomorphism between Octonionic Hilbert
Space (OHS) with complex geometry and the standard complex 
Hilbert space carrying the self-representation of the group 
$U(4)$. The octonionic structure is appropriately 
defined allowing a consistent tensor product of OHS.

In summary, life is not easy over generic OHS, we  cannot
define in a concise way tensor product and there is no notion of
hermiticity. But using complex geometry, all these problems soon vanish 
and our OHS behaves in suitable way.

We now give two examples of places where the nonassociativity of octonionic 
number is overcome by adopting complex scalar products.  A first objection, 
to the nonextendability to an octonionic formulation for the Hilbert space, 
concerns the completeness relation~\cite[pag.~50]{adl}. In writing the 
completeness formula  
\bea <\psi \mid \phi>  &=&   <\psi \mid \sum_{l} 
\Biggl( \mid \eta_{l}> <\eta_{l} \mid \Biggr) \mid \phi>  
\nonumber \\
&=& \sum_{l} <\psi \mid  \eta_{l}> \; <\eta_{l} \mid \phi> \q , 
\eea
we have assumed, in moving parentheses, that product of three factors is 
independent of the order of multiplication. If we take a two-dimensional 
Hilbert space with the following orthonormal states  
\bea
\mid \psi > ~ = \frac{1}{\sqrt{2}} \left(
\begin{array}{c}
e_4 \\
e_6 \end{array} \right) \quad \quad \mbox{and} \q \q 
\mid \phi > \; = \frac{1}{\sqrt{2}} \left(
\begin{array}{c}
-e_1 \\
e_3 \end{array} \right) \quad ,
\eea
\[ <\psi \mid \phi > ~ = 0.
\]
and for the complete set $ \mid \eta_l >$ we consider
\bea
\mid \eta_1 > \; = \left( \begin{array}{c} 
e_3 \\
0 \end{array} \right) \q \q \mbox{and} \quad \quad
\mid \eta_2 > \; =  \left( \begin{array}{c}
0 \\
e_5 \end{array} \right) \quad ,
\eea
we find 
\bea
\mid \eta_1> < \eta_1 \mid + \mid \eta_2 > < \eta_2 \mid
~ = \left( 
\begin{array}{cc} 
1 & 0 \\
0 & 1 \end{array} \right) = \openone_{2\times 2} \quad ,
\eea
and so 
\beal{w1}
<\psi \mid \Biggl( \sum_{l}
\mid \eta_{l} > < \eta_{l} \mid \Biggr) \mid \phi >  ~ = 0 \q .
\eea

Now, by direct calculations and  using the octonionic algebra, we have
\beal{w2}
\sum_{l} < \psi \mid \eta_{l} > < \eta_{l} \mid
\phi > ~ = e_5 \neq 0.
\eea

Nevertheless by requiring a complex projection for our scalar products 
\be
< ~~ \mid ~~ >_c ~~ \equiv ~~ 
\frac{1- e_1 \mid e_1}{2}~~< ~~ \mid ~~ > \q ,
\ee
eq.~(\ref{w1}) is obviously not changed, whereas eq.~(\ref{w2}) vanishes. 
So, by imposing complex geometry, we satisfy the completeness relation. 

A second objection in writing an octonionic quantum mechanics is the 
violation of the norm conservation~\cite[pag.~51]{adl}.
\bea
\dv_t <\psi(t) \mid \phi(t)> & = & \Biggl( \dv_t <\psi(t) \mid \Biggr)
\mid \phi(t) > +  <\psi(t) \mid  \Biggl( \dv_t \mid \phi(t) > 
\Biggr) \nonumber \\
& = & \Biggl( -<\psi(t) \mid \tilde{H}^{\dag}(t)\Biggr) \mid \phi(t) > +
<\psi(t) \mid \Biggl(- \tilde{H}(t) \mid \phi(t) >\Biggr) \nonumber \\
& = &
\Biggl( < \psi(t) \mid \tilde{H}(t) \Biggr) \mid \phi(t) >  + 
<\psi(t) \mid  \Biggl( - \tilde{H}(t) \mid \phi(t) > \Biggr).
\eea
where $\tilde{H}$ represents the anti-self-adjoint Hamiltonian. Again, 
working in a  two-dimensional Hilbert space, let 
$\mid \psi(0) >$ and $\mid \phi(0) >$ be orthonormal states
\bea
\mid \psi(0) > ~ = \frac{1}{\sqrt{2}} \left(
\begin{array}{c}
e_4 \\
e_6 \end{array} \right) \quad \quad  , \q \q 
\mid \phi(0) > ~ = \frac{1}{\sqrt{2}} \left(
\begin{array}{c}
-e_1 \\
e_3 \end{array} \right) \quad ,
\eea 
\[ <\psi(0) \mid \phi(0) > ~ = 0 \quad ,
\]
and consider
\bea
\tilde{H} ~ = ~ \left( \begin{array}{cc}
e_3 & 0 \\
0  & e_5 \end{array} \right) \quad .
\eea
The  time evolution operator will have the following form
\bea
U(t,0) = 
\left( \begin{array}{cc}
e^{-e_3 t} & 0 \\
0 & e^{-e_5 t} \end{array} \right)
= \left( \begin{array}{cc}
\cos t - e_3 \sin t & 0 \\
0 & \cos t - e_5 \sin t \end{array} \right) \quad ,
\eea
then we find
\bea
\mid \psi(t) > & = & U(t,0) \mid \psi(0) > ~ = ~
\frac{1}{\sqrt{2}} \left( \begin{array}{c}
e_4 \cos t - e_7 \sin t  \\
e_6 \cos t + e_3 \sin t \end{array} \right) \quad ,
\nonumber \\
\mid \phi(t) > & = & U(t,0) \mid \phi(0) > ~ = ~ 
\frac{1}{\sqrt{2}} \left( \begin{array}{c}
-e_1 \cos t + e_2 \sin t  \\
e_3 \cos t - e_6 \sin t \end{array} \right) \quad ,
\eea
leading to
\be
< \psi(t) \mid \phi(t) > ~ = ~ e_6 \sin t \cos t +
e_5 \sin^2 t \neq 0.
\ee
It is clear that by imposing complex scalar 
products, we have
\be 
< \psi(t) \mid \phi(t) >_c ~~= ~~ < \psi(0) \mid \phi(0) >_c ~~ = ~~ 0 \quad ,
\ee
which implies the norm conservation.  Everything is correct as it 
should be as required by the Rembieli\'nski's isomorphism (scalar 
product projects the OHS into the physical CHS).

We observe that the dimensionality of a complete set of states for complex 
inner product $<\psi \mid \phi>_{c}$ is {\em four times} 
that for the octonionic inner product $<\psi \mid \phi>$. 
Specifically if $\mid \eta_{l}>$ are a complete set of intermediate states for 
the octonionic inner product, so that
\[ <\psi \mid \phi> \; = \sum_{l} <\psi \mid \eta_{l}><\eta_{l} \mid \phi> \; 
\; ,\]
$\mid \eta_{l}>$, $\mid \eta_{l} \; e_{2}>$, $\mid \eta_{l} \;  e_{4}>$, 
$\mid \eta_{l} \; e_{6}>$ form a complete set of states for the complex 
inner product,
\bean 
\mid \phi> & = & \sum_{l} \;  ( \; 
\mid \eta_{l}><\eta_{l} \mid \phi>_{c}+ 
\mid \eta_{l} \; e_{2} ><\eta_{l} \; e_{2} \mid \phi>_{c}+\\
           &   & \; \; \; \; + 
\mid \eta_{l} \; e_{4}><\eta_{l} \; e_{4} \mid \phi>_{c}+ 
\mid \eta_{l} \; e_{6}><\eta_{l} \; e_{6} \mid \phi>_{c} \; ) \\
& = & \sum_{m} \mid \chi_{m}><\chi_{m} \mid \phi>_{c} \; \; ,
\eean
where $\chi_{m}$ represent {\em complex} orthogonal states. Thus the  
completeness relation can be written as 
\bean  
\stackrel{\raw}{\bf 1} & = & \sum_{m} \mid \chi_{m}>~~ << \chi_{m} \mid \; \; ,\\
\stackrel{\law}{\bf 1} & = 
& \sum_{m} \mid \chi_{m} >> ~~<\chi_{m} \mid 
\eean
(for further details on the 
completeness relation, one can consult an interesting work of Horwitz and 
Biedenharn~\cite[pag.~455]{hor}). So in our formalism we generalize the 
Dirac's notation by the definitions
\bean <<\chi_{m} \mid \phi> & = & 
<\chi_{m} \mid \phi>_{c} \; \; ,\\
<\phi \mid \chi_{m}>> & = & 
<\phi \mid \chi_{m} >_{c} \; \; .
\eean

\con

This paper aimed to give a clear exposition of the potentiality of 
barred octonions in quantum mechanics. We know that quantum mechanics is 
the basic tool for different physical applications. Many physicists believe 
that imaginary numbers are related to the deep secret of 
quantization. Penrose~\cite{penr} thinks that the quantization is completely 
based on complex numbers. Trying to overcome the problem of quantum gravity, 
he proposed to complexify the Minkowskian space-time. This represents the 
main assumption behind the twistor program. Adler~\cite{adl} believes that 
quantization processes should not be limited to complex numbers but should be 
extended to another member of the division algebras rank, the quaternionic 
field. He postulates that a successful unification of the fundamental 
forces will require a generalization beyond complex quantum mechanics. 
Adler envisages a two-level correspondence principle:
\bc
\bt{ccl}
 $\vert $       &                  & ~~~~~~classical physics and fields ,\\
 $\vert $       & ~distance scale  & ~~~~~~complex quantum mechanics and fields ,\\
 $\downarrow$  &                  & ~~~~~~quaternionic quantum field 
                                    dynamics (preonic level~\cite{adl1}) ,
\et
\ec
with quaternionic quantum dynamics interfacing with complex quantum theory, 
and then with complex quantum theory interfacing in the familiar manner 
with classical physics~\cite[pag.~498]{adl}.

Following this approach, we are tempted to postulate that octonionic quantum 
field theory may play an essential role in an even deeper fundamental level 
of physical structure.

Quaternionic quantum mechanics, using complex geometry~\cite{deleo1,deleo2} 
or quaternionic geometry~\cite{adl,adl1,adl2}, seems to be consistent from the 
mathematical point of view. Due to the octonions nonassociativity property, 
octonionic quantum mechanics seems to be a puzzle. In this paper we 
have presented an alternative way to look at the octonionic world. 

The first motivation, in using octonions numbers in physics, can be concisely 
resumed as follows: We hope to get a better understanding of standard 
theories if we have more than one concrete realization. In this way we can 
recognize the fundamental postulates which hold for any generic numerical 
field.

We emphasize that a characteristic of our formalism is the 
{\tt absolute need of a complex scalar product}, whereas in quaternionic 
quantum mechanics the use of a 
complex geometry is not obligatory and thus a question of choice~\cite{adl}. 
Using complex geometry we overcame the hermiticity problem and gave the 
appropriate and unique definition of momentum operator. We also realized 
Rembieli\'nski isomorphism and were able to write down a correct 
completeness relation and norm conservation. 

Obviously to check the consistence of our octonionic formulation of quantum 
mechanics we need an explicit application. So in a forthcoming paper 
we will develop a relativistic wave equation on the octonionic 
field~\cite{dekh}. 

We hope that the work presented in this paper, demonstrates that octonionic 
quantum mechanics may constitute a coherent and well-defined branch of 
theoretical physics. We are convinced that octonionic quantum mechanics 
represents largely uncharted and potentially very interesting, terrain in 
theoretical physics. 

\ack

The authors would like to thank P.~Rotelli for his important suggestions 
and for stimulating conversations. One of us (K.~A.~K.) gratefully 
acknowledges the warm hospitality of the Lecce Physics Department, where this 
paper was prepared.

\bref
\bi{jor}
P.~Jordan, Z.~Phys. {\bf 80}, 285 (1933).
\bi{wig}
P.~Jordan and E.~P.~Wigner, Ann.~Math. {\bf 35}, 29 (1934).
\bi{alb1}
A.~A.~Albert, Ann.~Math. {\bf 35}, 65 (1934).
\bi{fink1}
D.~Finkelstein {\em et al.}, {\em Notes on Quaternion Quantum Mechanics} 
(CERN, Report 59-7), in C.~Hoker, ed.~Logico-Algebraic-Approach to Quantum 
Mechanics II (Reidel, Dordrecht, 1979).
\bi{fink2}
D.~Finkelstein {\em et al.}, J.~Math.~Phys. {\bf 3}, 207 (1962); 
{\em ibid.} {\bf 4}, 788 (1963). 
\bi{gur2}
M.~G\"unaydin and F.~G\"ursey, \jxe{14}{1651}{73}; 
Lett.~Nuovo Cimento {\bf 6}, 401 (1973);  \pxf{9}{3387}{74}.
\bi{hor}
L.~P.~Horwitz and L.~C.~Biedenharn, \axp{157}{432}{84}. 
\bi{rem}
J.~Rembieli\'nski, \jxg{11}{2323}{78}.
\bi{sch}
R.~D.~Schafer,  {\it An introduction to Non-Associative Algebras} 
(Academic Press, New York, 1966).
\bi{oku}
S.~Okubo, {\it Introduction to Octonion and Other Non-Associative Algebras 
in Physics} (Cambridge University Press, Cambridge, to be published).
\bi{deleo1}
S.~De Leo and P.~Rotelli, \pxf{45}{575}{92}; \nxd{B110}{33}{95}; 
\pxxa{92}{917}{94}; \ixa{10}{4359}{95}; Mod.~Phys.~Lett.~A {\bf 11}, 357 
(1996).
\bi{deleo2}
S.~De Leo \pxxa{94}{11}{95};\\
S.~De Leo, {\em Quaternions and Special Relativity} 
(to be published in J.~Math.~Phys.);\\
S.~De Leo and P.~Rotelli, {\em Odd Dimensional Translation between Complex 
and Quaternionic Quantum Mechanics}  
(to be published in Prog.~Theor.~Phys.).
\bi{dekh}
S.~De Leo and K.~Abdel-Khalek, {\em Octonionic Dirac Equation}, 
Prog. Theor. Phys. (submitted).
\bi{adl}
S.~L.~Adler, {\it Quaternionic Quantum Mechanics and Quantum Fields} 
(Oxford, New York, 1995).
\bi{penr}
R.~Penrose and W.~Rindler, {\it Spinors and Space-Time} 
(Cambridge UP, Cambridge, 1984).
\bi{adl1}
S.~L~.Adler, \pxi{332B}{358}{94}.
\bi{adl2}
S.~L.~Adler,
 Phys.~Rev.~D {\bf 21}, 550 (1980);  
{\em ibid.} {\bf 21}, 2903 (1980); Phys.~Rev.~Lett. {\bf 55}, 783 (1985);  
Comm.~Math.~Phys. {\bf 104}, 611 (1986);  
Phys.~Rev.~D {\bf 34}, 1871 (1986);  
{\em ibid.} {\bf 37}, 3654 (1986); 
Phys.~Lett. {\bf 221B}, 39 (1989); 
\nxb{B415}{195}{94}.\\
A.~J.~Davies, Phys.~Rev.~D {\bf 41}, 2628 (1990).\\
C.~G.~Nash and  G.~C.~Joshi, J.~Math.~Phys. {\bf 28}, 463 (1987); 
Int.~J.~Theor.~Phys. {\bf 27}, 409 (1988).\\
L.~P.~Horwitz, J.~Math.~Phys. {\bf 34}, 3405 (1993); 
{\em ibid.} {\bf 35}, 2743 (1994); {\em ibid.} {\bf 35}, 2761 (1994).
\eref

\end{document}